\documentclass[11pt,a4paper]{article}

\usepackage{amsfonts}
\usepackage[dvips]{graphicx}
\usepackage{amsthm}
\usepackage{amsmath}
\usepackage{epsfig}
\usepackage{t1enc}
\usepackage{amssymb}
\usepackage{latexsym}
\usepackage{bm}

\def\F{{\cal F}}
\def\P{\Phi}
\def\t{\theta}
\def\ta{\tan \t}
\def\T{\Theta}
\def\tt{\dot \t}

\def\Ft{\dot \F}
\def\ttt{\ddot \t}

\def\Dt{\delta t}
\def\Dr{\delta r}
\def\l{\left(}
\def\r{\right)}
\def\L{\left[}
\def\R{\right]}

\def\e{\epsilon}
\def\a{\alpha}
\def\b{\beta}
\def\bt{\dot \beta}
\def\g{\gamma}
\def\o{\omega}

\title{\bf Evolution of a black hole-inhabited brane
\\
close to reconnection}

\author{Vladim\'ir Balek${}^1$\footnote{e-mail
address: balek@fmph.uniba.sk}\ \ and Branislav
Novotn\'y${}^2$\footnote{e-mail address: novotny@mat.savba.sk}
\\
${}^1${\it Department of Theoretical Physics, Comenius University,
Bratislava, Slovakia}
\\
${}^2${\it Mathematical Institute, Slovak Academy of Sciences,
Bratislava, Slovakia}}

\begin{document}

\renewcommand{\figurename}{Fig.}

\maketitle
\maketitle\abstract

{Last moments of a mini black hole escaping from a brane are
studied. It is argued that at the point of reconnection, where the
piece of the brane attached to the black hole separates from the
rest, the worldsheet of the brane becomes isotropic (light-like).
The degenerate mode of evolution, with the worldsheet isotropic
everywhere, is investigated. In particular, it is shown that the
brane approaches the reconnection point from below if it
reconnects within a certain limit distance, and from above if it
reconnects beyond that distance. The rate of relaxation to the
degenerate mode is established. If the dimension of the brane is
$p$, the nondegeneracy, measured by the determinant of the
relevant part of the induced metric tensor, falls down as
(latitudinal angle)$^{2(p - 1)}$.}

\section{Introduction}

One of the predictions of the theories with large extra dimensions
\cite{ark} is that there can exist mini black holes with masses of
order 1 TeV that can in principle be observed at LHC \cite{dim}.
In \cite{fro} it was pointed out that if a mini black hole has
been produced in a collision of two high-energy particles, it can
escape from the brane (that is, from our universe) after emitting
a hard quantum of Hawking radiation into the bulk. The description
of the brane-black hole system in \cite{fro} was quantum
mechanical. A classical description, which appeared first in
\cite{fro1} in the context of the problem of a static domain wall
interacting with a Schwarzschild black hole in 3 dimensions, was
applied to the escape of the black hole from the brane in
\cite{{fla},{fla1}}. In \cite{fla} it was established, by solving
numerically the equation of motion of the brane in the field of
the black hole, that the brane develops a neck that eventually
shrinks to a point. Then the piece of the brane attached to the
black hole cuts loose and the rest of the brane reconnects. In
\cite{fla1} it was argued that the scenario applies to the brane
with codimension one no matter what the velocity of the black
hole, but for greater codimensions the velocity must exceed some
critical value. Here we continue this study. In section
\ref{sec:eq} we write down the equation of motion of the brane, in
section \ref{sec:deg} we inspect the behavior of a special class
of solutions, which we call degenerate, close to reconnection, in
section \ref{sec:true} we investigate the relation between
degenerate and true solutions and in section \ref{sec:con} we
discuss the results.

%KAPITOLA 2

\section{Equation of motion} \label{sec:eq}

Consider a static $n$-dimensional black hole with a planar
$p$-dimensional brane attached to a given great sphere of the
horizon, and suppose that the black hole is knocked out of the
brane perpendicularly to it. The metric of an isolated black hole
is \cite{tan}
\begin{equation}
ds^2 = - fdt^2 + f^{-1} dr^2 + r^2 d\Omega^2, \quad f = 1 - r^{2 -
n}, \label{eq:metric}
\end{equation}
where $d\Omega^2$ is the metric of a unit $(n - 1)$-dimensional
sphere and we use a system of units in which $c = 1$ and $r_S$
(the Schwarzschild radius) $= 1$. The metric (\ref{eq:metric}) is
applicable, strictly speaking, only if the space the black hole is
living in is infinite and asymptotically flat, but can be used
also if some dimensions are compactified or warped, provided the
size of the black hole is much less than the scale on which that
happens. To describe the motion of the brane we pass to the rest
frame of the black hole and suppose that the brane has no
gravitational field of its own. Then the bulk space has metric
(\ref{eq:metric}). Furthermore, we assume that the brane is
infinitely thin. The dynamics of the brane is then governed by the
Dirac--Nambu--Goto action $S = - T$ $\times$ the volume of the
worldsheet, where $T$ is the tension of the brane and the volume
is computed from the metric of the bulk space induced on the
worldsheet of the brane. Finally we use the symmetry of the
problem. Let $\t$ be the latitudinal angle measured from the axis
that points opposite to the direction in which the black hole is
knocked out. The brane is obviously symmetric with respect to that
axis, therefore its worldsheet can be described, at least locally,
by the function $\t (r, t)$.

After computing the induced metric, inserting it into the
definition of the volume and integrating over the angles, we find
\cite{fla}
\begin{equation}
S = - T o \int \sqrt{F} (r\sin \theta)^{p - 1} drdt, \quad F = 1 +
r^2 f \t'^2 - r^2 f^{-1} \tt^2, \label{eq:action}
\end{equation}
where $o$ is the volume of a unit $(p - 1)$-dimensional sphere,
the prime denotes differentiation with respect to $r$ and the
overdot denotes differentiation with respect to $t$. Denote
$$q =2(p - 1), \quad h = r^{-2}f, \quad \F = hF = h + f^2 \t'^2 - \tt^2.$$
By varying $S$, we obtain the equation of motion of the brane
\begin{equation}
\P \equiv \ta (\tt \Ft - f^2 \t' \F' - 2 \T \F) - qh\F = 0,
\label{eq:eom}
\end{equation}
where
\begin{equation}
\T = \ttt - f^2 \t'' - f^2 {\cal R} \t', \quad {\cal R} = \frac pr
+ \frac {3f'}{2f}. \label{eq:dfT}
\end{equation}
The equation is linear in second derivatives of $\t$, with the
coefficients proportional to the components of the contravariant
metric tensor on the worldsheet. If the brane is to describe our
universe, its worldsheet has to be {\it timelike} (lay within the
lightcones) everywhere, with a possible exception of the point of
reconnection. For such worldsheet, the signature of the relevant
part of the worldsheet metric tensor is $(- +)$, hence the
equation is hyperbolic.

The character of the worldsheet is given by the sign of the the
determinant of the worldsheet metric tensor: the worldsheet is
timelike if the determinant is negative, isotropic if it is zero
and spacelike if it is positive. The determinant is proportional
to $- F = - \F/h$, therefore the three cases mentioned above
correspond to $\F > 0$, $\F = 0$ and $\F < 0$ respectively.
Reconnection of the brane takes place on the upper part of the
axis of symmetry of the brane (the half-line $\theta = 0$), where
the full determinant is zero since it contains an extra factor
$(r\sin \theta)^q$. However, this is just a coordinate effect and
the character of the worldsheet is again given by the sign of
$\F$. To see that, note that the character of the worldsheet is
complementary to that of the normal vector $n = \nabla [\t(r, t) -
\t]$ (the worldsheet is timelike if $n$ is spacelike, isotropic if
$n$ is isotropic and spacelike if $n$ is timelike), and the square
of the normal vector equals $r^{-2} F$.

The behavior of the brane at reconnection depends on whether it is
a string ($p = 1$) or a higher dimensional brane ($p > 1$). For a
string, the second term in equation (\ref{eq:eom}) is absent, and
the expression in the brackets in the first term is zero
everywhere including the reconnection point. In this case the
worldsheet stays timelike at reconnection. For higher dimensional
branes, if the expression in the brackets does not diverge at
reconnection, $\F$ must vanish there. As a result, the worldsheet
becomes {\it isotropic} (tangential to the lightcone). We will
show that this indeed happens if the function $\t$ satisfies some
natural requirements when approaching zero.

Rewrite $\P$ so that all derivatives $\ttt$ are absorbed into
$\Ft$ and all remaining derivatives $\tt'$ are absorbed into
$\F'$. In this way we obtain
\begin{equation}
\P \propto \ta (a \tt \Ft + f^2 b \t' \F' + 2 f^2 \Xi \F) - qh
\tt^2 \F, \label{eq:Prewrit}
\end{equation}
where
\begin{equation}
a = h + f^2 \t'^2, \quad b = \F - \tt^2, \quad \Xi =  \tt^2 (\t''
+ {\cal R} \t') - f^2 \t'^2 (\t'' + {\cal R}_0 \t') - \frac 12 h'
\t', \quad {\cal R}_0 = \frac {f'}f. \label{eq:dfabT}
\end{equation}
Denote by $t_0$ the time of reconnection and by $r_0$ the radial
coordinate of the point on the axis at which the reconnection
takes place. We are interested in the behavior of $\F$ near the
point $(r_0, t_0)$. Consider a function $\t$ that is smooth in the
vicinity of the reconnection point except possibly at this point
itself, and suppose the first two derivatives of $\t$ with respect
to $r$ are well-behaved at reconnection, $\t' \to 0$ and $\t''
\to$ positive number at $(r_0, t_0)$. Suppose furthermore that
$\t$ falls to zero linearly with $t$, $\tt \to$ negative number at
$(r_0, t_0)$. (Note that $\tt$ is necessarily finite at
reconnection since the function $\F$, which is by assumption
positive outside the point $(r_0, t_0)$, contains the term
$-\tt^2$. We have added only the requirements that $\tt$ has a
limit at $(r_0, t_0)$ and that that limit is nonzero. Note also
that the assumption about nonzero $\tt$ close to reconnection has
been already used in (\ref{eq:Prewrit}), where we have suppressed
the general factor $\tt^{-2}$ on the right hand side.) The
assumptions seem plausible, if for nothing else because of the
shape of the curves in \cite{fla}. A principal consequence is that
$\F$ has a nonnegative limit at $(r_0, t_0)$, so that all we have
to show is that the limit is zero. Our starting point will be
equation $\P = 0$ with $\P$ given in (\ref{eq:Prewrit}), regarded
as a first order differential equation for $\F$. The equation is
quasilinear and can be solved by the method of characteristics
\cite{arn}. For the three variables $t$, $r$ and $\F$ as functions
of $\lambda$ we have three ordinary differential equations of
first order,
\begin{equation}
d_\lambda t = a\tt, \quad d_\lambda r = f^2 b \t', \quad d_\lambda
\F = - (2 \Xi - q h \cot \t \tt^2) \F, \label{eq:charF}
\end{equation}
where $d_\lambda$ denotes derivative with respect to $\lambda$.
The curves in the 3-dimensional space $(r, t,\F)$ defined by these
equations are called {\it characteristics} of equation $\P = 0$.
Close to the point $(r_0, t_0)$ we have
$$d_\lambda t \sim h_0 \t_{10}, \quad d_\lambda r \sim  0, \quad d_\lambda
\F \sim q h_0 \t_{10}^2 \t^{-1} \F,$$ where $h_0$ and $\t_{10}$
are the values of $h$ and $\tt$ at reconnection. Denote $\Dt = t -
t_0$. For the characteristic that reaches the reconnection point
we have $\t \sim \t_{10} \Dt$, and after inserting this into the
third equation and dividing the third equation by the first, we
obtain $d_t \F \sim q\F/\Dt$ and $\F \sim$ {\it const} $(-
\Dt)^q$. The proof is completed.

To show that the worldsheet becomes isotropic at reconnection we
needed some {\it ad hoc} assumptions about its form. Also, we did
not address the question whether the worldsheet stays timelike
{\it outside} the reconnection point. The assumptions can be
relaxed and the missing proof supplied with the help of a
conservation law known from the string theory. Let us start with
an observation that a higher dimensional brane can be replaced by
a string living in the 2 + 1 dimensional space $(r, \t, t)$, with
an effective metric $(r\sin \theta)^{q/2}$ $\times$ the true
metric of the space $(r, \t, t)$ (equal to $- fdt^2 + f^{-1} dr^2
+ r^2 d\t^2$). In a static gravitational field, it is convenient
to introduce static gauge with the coordinate $\sigma$ numerating
the points of the string chosen in such a way that the string
moves perpendicularly to itself. The energy stored in
infinitesimal segments of the string is then conserved not only in
a sum, but also separately. The energy is proportional to $\sqrt{-
g_{00}}\ dl/\sqrt{1 - v^2}$, where $dl$ is the length of the
segment and $v$ is its velocity. (For the derivation of the
formula in flat space, see \cite{zwi}.) The worldsheet is timelike
if $v < 1$, so that the strip of the worldsheet corresponding to
the given segment of the string remains timelike for all times if
it was timelike at the beginning. This holds unless the string
develops {\it cusps}, where $v \to 1$ simultaneously with $dl \to
0$ (the segment at the tip of the cusp is contracted to a point,
and moves perpendicularly to the cusp with the velocity of light).
A string in the effective metric we are interested in behaves in
the same way at reconnection: both $\sqrt{- g_{00, eff}}$ and
$dl_{eff}$ approach zero there because of the collapse of the
metric, therefore $v_{eff}$ must approach 1. However, the velocity
remains unchanged as we return from the effective metric to the
true one. As a result, the velocity $v$ of the brane must approach
1 as the brane moves towards the reconnection point.

In \cite{fla}, the initial velocity of the brane is chosen as $v =
(1 - 1/r) v_\infty$, where $v_\infty$ is the velocity of the brane
at infinity (the velocity with which the black hole is knocked out
of the brane with the sign minus), and the calculation is
performed for three different values of $v_\infty$ from the
interval $0 < v_\infty \le 1$. (The value $v_\infty = 1$ has been
obviously included into the list in order to cover the case
$v_\infty < 1$, $1 - v_\infty \ll 1$.) For such choice of $v$, $\F
= h(1 - v^2)$ is positive, and the worldsheet is timelike, at the
moment when the black hole receives the initial push that puts it
into motion. Consequently, if the brane does not develop cusps,
$\F$ stays positive and the worldsheet stays timelike up to the
point of reconnection, where $\F$ vanishes and the worldsheet
becomes isotropic.

%KAPITOLA 3

\section{Properties of degenerate solutions}  \label{sec:deg}

We are interested in the behavior of a brane that is just about to
reconnect. If the brane becomes isotropic at the point of
reconnection, it should be approximately isotropic close to that
point. Thus, it is natural to start with the brane that is {\it
exactly} isotropic. Such brane is described by the equation
\begin{equation}
\F \equiv h + f^2 \t'^2 - \tt^2 = 0. \label{eq:eom1}
\end{equation}
The equation can be viewed as an initial condition imposed on the
solutions of equation $\P = 0$. Indeed, since $\P$ is homogeneous
in $\F$, the function $\t$ that satisfies $\F = 0$ at a certain
moment, and is evolving according to $\P = 0$, will satisfy $\F =
0$ ever after. (The brane that was once isotropic remains always
isotropic.) On the other hand, one can view (\ref{eq:eom1}) as a
dynamic equation. Since it is of the first order only, we need
half as much initial data for it than for the equation $\P = 0$;
we get along with the function $\t$ and do not need to know the
function $\tt$ at the beginning. With regard to this, solutions to
$\P = 0$ that satisfy also $\F = 0$ can be called {\it
degenerate}.

To solve equation (\ref{eq:eom1}) close to reconnection, expand
$\t$ in the powers of $\Dt = t - t_0$ and $\Dr = r - r_0$,
\begin{equation}
\t = \t_{10} \Dt + \frac 12 \t_{20} \Dt^2 + \t_{11} \Dt \Dr +
\frac 12 \t_{02} \Dr^2 + \ldots \label{eq:expth}
\end{equation}
The coefficients $\t_{00}$ and $\t_{01}$ are missing because the
brane reconnects on the axis, and touches the axis while
reconnecting rather than intersects it. At the moment of
reconnection, $\t$ reduces to
$$\t_0 = \frac 12 \t_{02} \Dr^2 + \frac 16 \t_{03} \Dr^3 + \ldots,$$
where $\t_{02}$ is supposed to be nonzero and positive. The
coefficients $\t_{0n}$ entering this expression must be fixed in
advance. By choosing them, we provide equation (\ref{eq:eom1})
with an initial condition. (Except that we must evolve the
solution backwards in time, so that the condition is final rather
than initial.) All coefficients  $\t_{mn}$ with $m > 0$ can be
expressed in terms of the coefficients $\t_{0n}$ by solving
equation (\ref{eq:eom1}) order by order. The first few
coefficients are
\begin{equation}
\t_{10} = - \sqrt{h_0}, \quad \t_{20} = 0, \quad \t_{11} = \frac
{h_1}{2\t_{10}}, \quad \t_{12} = \frac {\frac 12 h_2 + f_0^2
\t_{02}^2 - \t_{11}^2} {\t_{10}}, \label{eq:coeff}
\end{equation}
where we have denoted the value of $f$ at $r = r_0$ by $f_0$ and
the value of the $n$th derivative of $h$ at $r = r_0$ by $h_n$.
The square root in the expression for $\t_{10}$ is taken with the
sign minus in order to ensure that $\t$ decreases before
reconnection. Of the three coefficients of third order, $\t_{30}$,
$\t_{21}$ and $\t_{12}$, we list only the last one since the other
two will not be needed in what follows.

Expansion of $\t$ can be used to determine, for example, in what
direction the brane approaches the reconnection point or by what
rate it is either broadening or narrowing in the process, given
the location of the reconnection point and the final shape of the
brane. In this way we learn how an isotropic brane can possibly
move just before reconnection. The resulting picture is
applicable, as we will see, to a timelike brane as well.

\begin{figure}[ht]
\centerline{\includegraphics[height=4cm]{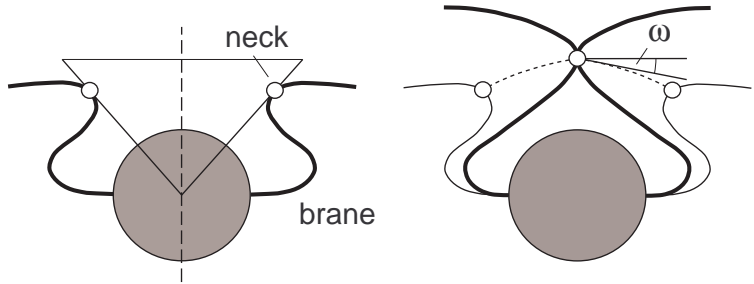}}
\caption{\small Motion of the brane close to reconnection
(schematically)} \label{fig:neck}
\end{figure}

Define the {\it neck} of the brane as the $(p - 1)$-dimensional
sphere at which the brane touches the cone $\t =$ {\it const}. The
location of the neck is shown in figure \ref{fig:neck} on the
left.
Define, furthermore, the angle $\o$ at which the brane approaches
the reconnection point as the angle between the direction in which
the neck arrives at the axis and the plane perpendicular to the
axis. This is depicted in figure \ref{fig:neck} on the right. If
we are interested in the behavior of the brane as visualized in
pictures like this, it is more instructive to compute the {\it
apparent} angle (the angle referring to the Euclidean geometry on
the plane) than the true one. Thus, we define
$$\tan \o =  - \left. \frac {\Dr_N}{r_N \t_N}\right|_{\Dt = 0},$$
where $r_N$ and $\t_N$ is the radial and latitudinal coordinate of
the neck. (The true value differs from this by the factor
$f_0^{-1/2}$.) The location of the neck is given by the condition
$$\t'\big|_{r = r_N} = 0.$$
From this we obtain $r_N$ as a function of $t$, and by inserting
it into $\t (r, t)$ we find $\t_N$ as a function of $t$. To
compute $\tan \o$, we need to know only the leading term in the
expansions of $\Dr_N$ and $\t_N$ in $\Dt$. By the procedure
described above we obtain
\begin{equation}
\Dr_N = - \frac {\t_{11}}{\t_{02}} \Dt, \quad \t_N = \t_{10} \Dt,
\label{eq:neck}
\end{equation}
and after inserting this into the definition of $\tan \o$ we find
\begin{equation}
\tan \o = \frac {\t_{11}}{r_0 \t_{10} \t_{02}}. \label{eq:al}
\end{equation}
With the help of (\ref{eq:coeff}), this can be rewritten as $\tan
\o$ = positive number $\times$ $h_1$ = positive number $\times$
$(-2 + nr _0^{-n + 2})$. As a result, the sign of $\o$ depends on
the radius $r_0$,
\begin{equation}
\o \mbox{\hskip 1mm} \bigg \{ \mbox{\hskip -2.5mm} \left.
\begin{array} {l}
  > 0\ \mbox{for}\ r_0 < r_{0crit}\\
  < 0\ \mbox{for}\ r_0 > r_{0crit}\\
  \end{array} \right. \mbox{\hskip -2mm},
\label{eq:al1}
\end{equation}
with the critical radius defined as
\begin{equation}
r_{0crit} = \l \frac n2\r^{\frac 1{n - 2}}.\label{eq:rcrit}
\end{equation}
The brane arrives at the axis from below if the reconnection takes
place inside the sphere with the radius $r_{0crit}$, and from
above if the reconnection takes place outside that sphere.

The shape of the brane close to reconnection can be characterized
by its apparent curvature at the neck
$$k = r_N \t''_N.$$
(The true value differs from this by the factor $f_N$.) Let us
determine how this quantity varies with time just before
reconnection. After inserting $\Dr = \Dr_N + \e$ into the
expansion of $\t$ and collecting the terms of zeroth and first
order in $\Dt$ that are proportional to $\e^2$, we find
$$\t = \t_N + \frac 12 (\t_{02} + \t_{12} \Dt + \t_{03} \Dr_N)\e^2,$$
so that
$$k = (r_0 + \Dr_N)(\t_{02} + \t_{12} \Dt + \t_{03} \Dr_N) =
k_0 + r_0 \t_{12} \Dt + (\t_{02} + r_0 \t_{03}) \Dr_N,$$ where
$k_0 = r_0 \t_{02}$ is the apparent curvature of the brane at the
point of reconnection. If we insert here $\Dr_N$ from the first
equation (\ref{eq:neck}), we obtain
$$k = k_0 + \L r_0 \t_{12} - \l 1 + r_0 \frac {\t_{03}}{\t_{02}}\r
\t_{11} \R \Dt.$$ Thus, the time derivative of the curvature at
the point of reconnection is
\begin{equation}
\dot k_0 = r_0 \t_{12} - \l 1 + r_0 \frac {\t_{03}}{\t_{02}}\r
\t_{11}. \label{eq:k}
\end{equation}

\begin{figure}[ht]
\centerline{\includegraphics[height=6.2cm]{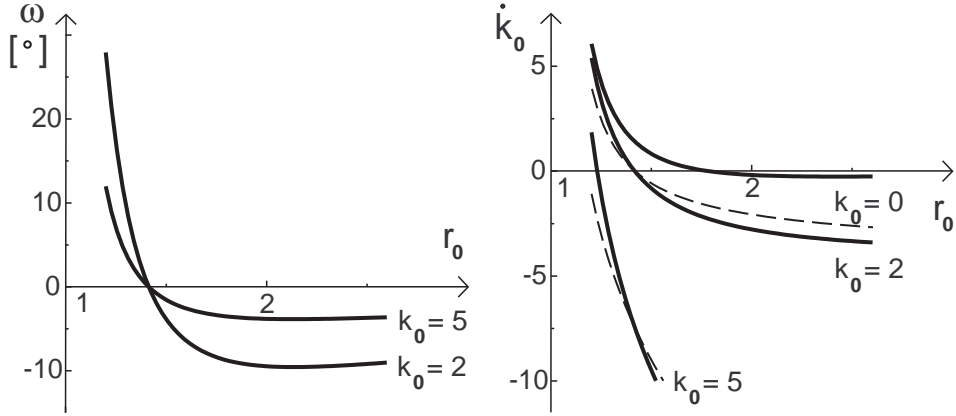}}
\caption{\small Parameters of the brane at reconnection}
\label{fig:par}
\end{figure}

In figure \ref{fig:par} we plot the angle $\o$ and the time
derivative of the curvature $\dot k_0$ as functions of the radius
$r_0$.
Both parameters depend on the curvature $k_0$, and $\dot k_0$
depends also on the parameter of asymmetry $l_0 = r_0 \t_{03}$. In
the right panel, solid lines correspond to $l_0 = 0$ and dashed
lines to $l_0 = -5$ for $k_0 = 2$ and $l_0 = -25$ for $k_0 = 5$.
The brane is supposed to live in a 4-dimensional space. (Its own
dimension does not affect its motion in case it is isotropic.) For
such brane, the critical radius is $r_{0crit} = \sqrt{2}$. Both
$\o$ and $\dot k_0$ decrease monotonically with $r_0$, but pass
through zero, in general, at different points: $\o$ at $r_{0crit}$
and $\dot k_0$ at a point depending on $k_0$ and $l_0$. In a space
with one extra dimension, the latter point is located at
$r_{0crit}$ for $k_0 = 2$. Thus, the brane that arrives at the
axis with the curvature 2 (in ordinary units $2/r_S$) is narrowing
just before touching the axis if the reconnection takes place
inside the sphere with the radius $r_{0crit}$, and broadening if
the reconnection takes place outside that sphere.

To see how an isotropic brane actually approaches the point of
reconnection, let us solve equation $\F = 0$ numerically. We can
use once again the method of characteristics, this time for a {\it
nonlinear} differential equation \cite{arn}. To simplify formulas,
replace the Schwarzschild coordinate $r$ by the ``tortoise
coordinate'' $R = \displaystyle \int f^{- 1} dr$. Then
\begin{equation}
\F = h + \hat \t^2 - \tt^2, \label{eq:FR}
\end{equation}
where the hat denotes differentiation with respect to $R$. We want
to solve equation $\F = 0$ with the initial condition $\t = \t_i$
at some moment $t_i$. Introduce an auxiliary Hamiltonian obtained
by replacing the derivatives $\hat \t$ and $\tt$ in $\F$ by the
momenta $\pi_R$ and $\pi_t$, and adding an extra factor $1/2$ for
convenience,
$$H = \frac 12 \l h + \pi_R^2 - \pi_t^2 \r.$$
The Hamiltonian lives in the phase space $x^A = (R, t)$, $\pi_A =
(\pi_R, \pi_t)$. The solution to equation $\F = 0$ is given by the
Hamilton equations for $R$ and $t$ and an additional equation for
$\t$ as functions of the ``time'' $\lambda$,
\begin{equation}
d_\lambda R = \pi_R, \quad d_\lambda \pi_R = - \frac 12 \hat h,
\quad d_\lambda t = - \pi_t, \quad d_\lambda \pi_t = 0, \quad
d_\lambda \t = \pi\ .\ d_\lambda x = \pi_R^2 - \pi_t^2,
\label{eq:char}
\end{equation}
with the initial conditions
\begin{equation}
R = R_*, \quad \pi_R = \hat \t_*, \quad t = t_i, \quad \pi_t =
\tt_* = - \sqrt{h_* + \hat \t_*^2}, \quad \t = \t_*.
\label{eq:chin}
\end{equation}
The star refers to the value of the function at $R_*$, or if it
depends on both $R$ and $t$, at $(R_*, t_i)$; hence $\t_* = \t_i
(R_*)$ and $\hat \t_* = \hat \t_i (R_*)$. The curves given by
equations (\ref{eq:char}) are characteristics of equation $\F =
0$. From the initial conditions it follows that $H = 0$ at the
starting point of the characteristic, and since $H$ does not
depend explicitly on $\lambda$, it is conserved and we have $H =
0$ along the whole characteristic. This allows us to simplify the
last equation in (\ref{eq:char}) to $d_\lambda \t = - h$. In
addition to that, since $\pi_t$ is constant, we could use equation
$H = 0$ to compute $\pi_R$. However, the resulting expression is
not very helpful in numerical calculations because of the
ambiguity of its sign, therefore it is preferable to compute
$\pi_R$ from the differential equation. By combining equations for
$R$, $\t$ and $\pi_R$ with that for $t$, and returning from $R$ to
$r$, we arrive at the equations
\begin{equation}
\frac {dr}{dt} = - \frac {f \pi_R}{\pi_t}, \quad \frac
{d\pi_R}{dt} = \frac{fh'}{2\pi_t}, \quad \frac {d\t}{dt} = \frac
h{\pi_t}. \label{eq:eqchart}
\end{equation}
The value of $\pi_t$ and the initial value of $\pi_R$ are given in
(\ref{eq:chin}), with $\hat \t_*$ replaced by $f_* \t_*'$.

The curves in the 2 + 1 dimensional space $(r, \t, t)$ we have
constructed are in fact null geodesics. It is seen most easily if
we multiply the $h$-term in the Hamiltonian by $\pi_\t^2$, since
then the Hamiltonian transforms into $H = \dfrac 12 g_{eff}^{\mu
\nu} \pi_\mu \pi_\nu$ with the effective metric $g_{eff}^{\mu \nu}
= f g^{\mu \nu}$. (If we are interested in {\it null} geodesics,
metric can be rescaled by an arbitrary function.) The previous
theory is completely reproduced if we add one more initial
condition $\pi_\t = -1$ to the conditions (\ref{eq:chin}). We can
also notice that equation $\F = 0$ is {\it eikonal equation} in
the space $(t, r, \t)$, with the eikonal defined as $\t (r, t) -
\t$. Thus, the worldsheet of the brane can be viewed as {\it wave}
(hypersurface of constant phase), and the curves we have
constructed as {\it rays} (curves that are at the same time normal
and tangential to the wave). The fact that rays are null geodesics
is a well-known result of geometrical optics in curved spacetime
\cite{mtw}.

\begin{figure}[ht]
\centerline{\includegraphics[height=6.3cm]{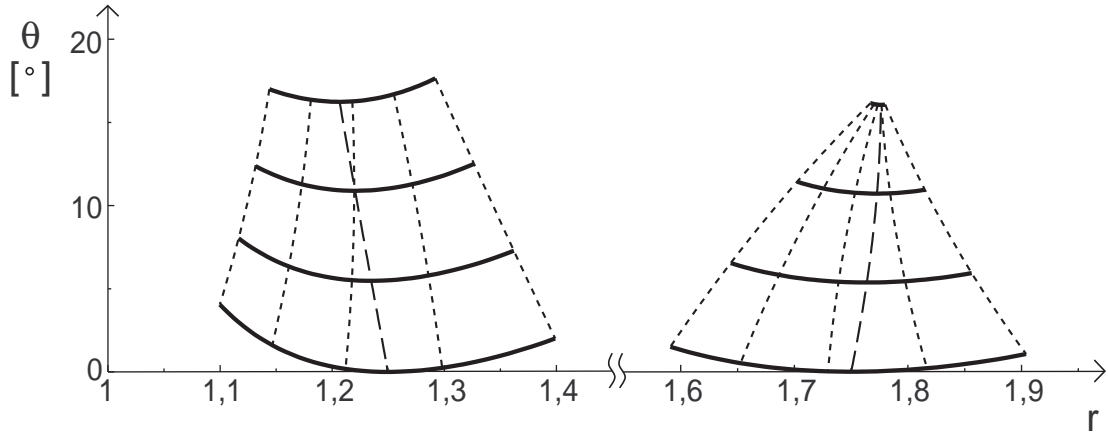}}
\caption{\small Degenerate evolution of the brane}
\label{fig:evol}
\end{figure}

In figure \ref{fig:evol} we depict two branes with parabolic shape
at the time 0.6 before reconnection, ending up
at the point $r_0 = 1.25$ (left panel) and $r_0 = 1.75$ (right
panel) with the curvature $k_0 = 5$ and 3 respectively. Both
branes are asymmetric at reconnection (have nonzero $l_0$). Heavy
lines are graphs of $\t(r, t)$ at different times, dotted lines
are characteristics used in the computation and dashed lines are
the paths of the neck of the brane. The brane on the left
approaches the axis from below while the brane on the right
approaches it from above, at the angle $\o$ equal to 7.1$^\circ$
and $-5.6^\circ$ respectively. The further from the horizon the
more divergent the characteristics, therefore the brane on the
right has smaller range of allowed curvatures at reconnection than
the brane on the left. For example, if we raised the curvature of
the former brane to $k_0 = 5$ and evolved it backwards in time,
the procedure would collapse at the time about 0.3 before
reconnection. The characteristics would start to intersect and the
curves would develop cusps. It must be stressed, however, that no
pathology arises if the brane moves {\it forward} in time. If an
inward bent, shrinking isotropic brane appears near the black
hole, it stays smooth until its neck shrinks to a point.

%KAPITOLA 4

\section{Relaxation of the true solution to the degenerate one}
\label{sec:true}

Let us turn to the brane whose worldsheet is timelike before
reconnection. We have seen that such brane becomes isotropic when
reconnecting; thus, $\F$ relaxes to zero as $\t$ approaches zero.
Now we address the question how {\it fast} it relaxes.

Let us write $\F$ for $\t \sim 0$ as $\F \sim a \t^Q$, where $\a$
is a function of $r$ and $Q$ is a positive constant. The
possibility to express $\F$ in this way is not self-evident; it is
rather an {\it assumption} which can be accepted only if we prove
that it is consistent with equation (\ref{eq:eom}). For $\F$ of
the given form, the left hand side of the equation becomes
$$\P \sim \tan \t \L \a Q \t^{Q - 1} \tt^2 - f^2 \l \a' \t^Q +
\a Q \t^{Q - 1} \t' \r \t' - 2\T \a \t^Q\R - qh\a \t^Q.$$ If we
keep here only the leading term, the expression simplifies to
$$\P \sim \a \t^Q [Q ( \tt^2 - f^2 \t'^2 ) - qh].$$
This is zero in the leading order if $Q = q$, since then the
expression in the square brackets equals $- q\F$ and is of order
$\t^q$. This suggests that $\F$ goes to zero with $\t$ as
\begin{equation}
\F \sim \a \t^q. \label{eq:F0}
\end{equation}
Denote the value of $\a$ at the reconnection point by $\a_0$. If
$\a_0$ is nonzero, we can write the relaxation law for $\F$ as $\F
\sim \a_0 \t^q$, or $\F \propto \t^q$. In section \ref{sec:eq} we
obtained the asymptotics $\F \propto (- \Dt)^q$, valid along a
curve approaching the reconnection point perpendicularly to the
axis (so that $\t \propto - \Dt$). The present formula generalizes
this asymptotics to an arbitrary direction.

To complete the discussion, let us return to the assumption that
$\a_0$ is nonzero. This is valid generically, and {\it must} be
valid if the brane develops no cusps during its evolution and
meets the requirements used in the derivation of the asymptotics
$\F \propto (- \Dt)^q$. To prove that, consider again the ordinary
differential equation for $\F$ in section \ref{sec:eq} (the last
equation in (\ref{eq:charF})). It is a homogeneous linear equation
of first order, hence its general solution can be written as $\F =
C \F_{ref}$, where $\F_{ref}$ is a reference solution and $C$ is
an arbitrary constant. Since $\F \propto (- \Dt)^q$ for $\Dt \sim
0$, the reference solution can be written as $\F_{ref} = (- \Dt)^q
+$ higher order terms in $\Dt$. If $C$ is zero, $\cal F$ is zero
all the way down the characteristic that terminates at the
reconnection point. However, in the nondegenerate mode of
evolution with no cusps $\F$ is positive everywhere outside the
reconnection point; and $\F_{ref}$ is positive at least in {\it
some} interval $\Dt < 0$. Thus, $C$ must be positive, and the
value of $\a$ at the reconnection point, equal to $C (-
\t_{10})^{-q}$, must be positive, too.

The asymptotics of $\F$ that we have found allows us to determine
how $\F$ depends on the variables $r$ and $t$, provided we know
the dependence of $\t$ on them. The simplest possibility is that
$\t$ is {\it regular} in the sense that it can be Taylor-expanded
in $\Dr$ and $\Dt$, as we have already assumed in the degenerate
case. Then $\F$ is regular, too, and its Taylor expansion begins
by the terms proportional to $\Dt^q$, $\Dt^{q - 1} \Dr^2$,
$\ldots$, $\Dr^{2q}$. However, equation for $\t$ is singular at
reconnection, because the coefficients at the highest order
derivatives are zero there. Thus, to decide whether $\t$ is
regular or not we need to analyze the effect of this singularity
on the character of the solution. In what follows we aim to show
that $\t$ {\it can} be regular; that is, that the assumption of
regularity of $\t$ does not lead to contradiction. For that
purpose we will combine the expansion of $\t$ in $\Dr$ and $\Dt$
with the expansion of $\F$ in $\t$. This turns out to be more
effective than to expand all quantities in $\Dr$ and $\Dt$ from
the very start.

Let us write $\F$ as a power series in $\t$,
\begin{equation}
\F = \a \t^q + \b \t^{q + 1} + \ldots, \label{eq:F}
\end{equation}
where $\a$ is as before a function of $r$ and the other
coefficients are functions of both $r$ and $t$. Since $\t$ is
itself a function of $r$ and $t$, the expansion is not unique.
However, for our purposes it is sufficient to have a {\it certain}
expansion of this form, as long as it is well-defined and
consistent with equation (\ref{eq:eom}). To obtain such expansion,
we start by replacing $h$ in the last term in $\P$ by ${\cal H} +
\F$, where ${\cal H} = \tt^2 - f^2 \t'^2$. Then we fix $\t$, so
that equation (\ref{eq:eom}) becomes a first order differential
equation for $\F$. Finally we insert the expansion of $\F$ into
(\ref{eq:eom}) and collect the terms containing $\t$ in the powers
$q$, $q + 1$, $q + 2$, $\ldots$ {\it explicitly}. In this way we
obtain a system of equations for the expansion coefficients. The
first equation is satisfied identically, so that we have no
restriction on $\a$. (This could be seen in advance from the
considerations at the beginning of this section.) The remaining
equations can be solved order by order to obtain expressions for
$\b$, $\g$, $\ldots$ in terms of $\a$. In particular, the
coefficient appearing in the next-to-leading order is
\begin{equation}
\b = \frac 1{\cal H} (f^2 \t' \a' + 2 \T \a + \a^2 \delta_{q1}).
\label{eq:beta}
\end{equation}
In this expression, the quadratic term is present only for the
nonphysical value $q = 1$ (``one-and-a-half dimensional brane'').
For an arbitrary $q$, the first $q - 1$ coefficients starting from
$\b$ are linear in $\a$, and a higher power of $\a$ appears first
in the $q$th coefficient, in front of $\t^{2q}$.

The previous procedure enables us to express $\F$ for a given $\t$
solely in terms of an arbitrary function of one variable $\a$.
This agrees with the fact that for a first order differential
equation we need one initial condition. With our expression for
$\F$, we can satisfy one constraint imposed on $\F$ by a proper
choice of $\a$. On the other hand, the possibility to fix the
initial condition in this way explains why we could have chosen
$\a$ as a function of $r$ only.

To find $\t$ as a function of $r$ and $t$, we must supplement
equation (\ref{eq:eom}) by the definition of $\F$. Thus, we must
solve {\it two} equations
\begin{equation}
\P = 0, \quad \Psi \equiv \tt^2 - f^2 \t'^2 - h + \F = 0.
\label{eq:PP}
\end{equation}
regarded as equations for $\F$ and $\t$ respectively. The
equations look as if they were of first order, but we must keep in
mind that $\P$ contains derivatives of $\t$ up to second order. If
we transformed $\P$ into the form (\ref{eq:Prewrit}), we would get
rid of the derivatives $\ttt$ and $\tt'$, but the global picture
would stay the same; therefore we stick to the simpler expression
(\ref{eq:eom}) for $\P$. The presence of the derivatives of $\t$
in $\P$ apparently ruins the idea of solving the equations by
expanding $\F$ in the powers of $\t$. If we expand $\F$ and solve
the first equation order by order, in the resulting expression
there will appear derivatives of $\t$ of all orders (second
derivatives in $\b$, third derivatives in $\g$ etc.). After
inserting this into the second equation we obtain an equation of
infinite order, which is unacceptable, if for nothing else because
we need infinitely many initial conditions to fix the solution.
Nevertheless, the equation {\it can} be solved by expanding $\t$
into the powers of $\Dr$ and $\Dt$. The point is that the
coefficients of higher order in $\Dt$ appearing in the expressions
for the coefficients of lower order in $\Dt$ can be calculated in
advance, since they have necessarily a lower {\it total} order
(sum of the orders in $\Dr$ and $\Dt$). In other words, algebraic
equations for the coefficients can be solved one by one if lined
up by their total order. The details can be found in appendix
\ref{2step}.

As noted before, the arbitrary function $\a$ appears in the
expansion of $\F$ because of the freedom of choice of the initial
condition for $\F$. Write $\F$ as $\F = \xi \t^q$. The initial (in
fact, final) conditions for both $\t$ and $\F$ can be imposed at
the moment of reconnection, by choosing $\t_0$ and $\F_0$, or
equivalently, $\t_0$ and $\xi_0$. Once $\t_0$ is chosen, $\a$ must
be given uniquely by $\xi_0$. In fact, we can obtain expansion
coefficients $\a_n$ of $\a$ by computing them order by order from
expansion coefficients $\xi_{0n}$ of $\xi_0$. For more detail, see
the second part of appendix \ref{2step}.

Once we have assumed that $\t$ can be expanded in $\Dr$ and $\Dt$,
we could have expanded both equations for $\F$ and $\t$ in $\Dr$
and $\Dt$ immediately, instead of expanding first the former
equation in $\t$. In such approach, we would expect that there
exist no equations for the expansion coefficients $\F_{qn}$. The
reason is that $\F_{qn}$ are in one-to-one correspondence with
$\xi_{0n}$, which in turn are in one-to-one correspondence with
$\a_n$; and since $\a_n$ are free, $\F_{qn}$ must be free, too.
However, after actually expanding equations for $\F$ and $\t$ in
$\Dr$ and $\Dt$, we obtain an infinite string of equations for
$\F_{qn}$. The fact that $\F_{qn}$ are free leads to the
conclusion that, after all $\t$'s appearing in the equations are
expressed in terms of $\t_{0n}$ and $\F_{qn}$, the equations must
collapse to $0 = 0$ due to massive cancelations. The argument is
elaborated in appendix \ref{1step}.

To summarize the previous discussion, if $\t$ expands in $\Dr$ and
$\Dt$, we can compute its coefficients in terms of expansion
coefficients of $\t_0$ and $\xi_0$; however, as for now we cannot
tell whether the resulting series converges or not. To obtain a
tentative answer, consider equation (\ref{eq:eom}) with the
initial conditions $\t = \t_0$ and $\xi = \xi_0$ {\it outside} the
reconnection point. If the initial conditions are formulated on
two intervals $I_-$ and $I_+$ on the axis $r$ to the left and to
the right of the point $r = r_0$, the equation can be solved
inside two strips ${\cal I}_-$ and ${\cal I}_+$ in the plane $(r,
t)$, bounded from above by the intervals $I_-$ and $I_+$ shifted
to the line $t = t_0$, and from the sides by the diverging
characteristics of equation (\ref{eq:eom}). If, furthermore, the
initial conditions are smooth, the solution must be smooth, too.
For $\t_0$ and $\xi_0$ that can be both expanded into Taylor
series around the point $r = r_0$ on an interval $I$ containing
$I_-$ and $I_+$, this implies that $\t$ is smooth in the domains
${\cal I}_-$ and ${\cal I}_+$. On the other hand, $\t$ can be
written as the Taylor series we have formally introduced earlier,
hence the series must be convergent in both domains. If so, it
seems plausible that it is convergent in the strip between them,
too.

The first nonzero coefficient $\F_{mn}$ is $\F_{q0}$, followed by
$\F_{q - 1,2}$, $\F_{q,1}$, $\F_{q + 1, 0}$ etc.; thus, the first
coefficient $\t_{mn}$ affected by nondegeneracy is $\t_{q + 1,
0}$, followed by $\t_{q2}$, $\t_{q + 1,1}$, $\t_{q + 2, 0}$ etc.
In particular, for the nonphysical value $q = 1$ equation $\t_{20}
= 0$ does not hold any longer, but is replaced by $\t_{20} = -
\F_{10}/(2\t_{10})$. Also, in $\t_{12}$ there appears a new term
$- \F_{02}/(2\t_{10})$ with $\F_{02} = \t_{02} \F_{10}/\t_{10}$.
All $\t_{mn}$ of lower order than those cited above are the same
as for an isotropic brane. Because of that, equations
(\ref{eq:al}) and (\ref{eq:k}) for $\a$ and $\dot k_0$, derived
for degenerate evolution, stay valid for nondegenerate evolution
in almost all cases. The only exception is the second equation in
case $q = 1$, which is modified because of the additional term in
$\t_{12}$.

\begin{figure}[ht]
\centerline{\includegraphics[height=6.3cm]{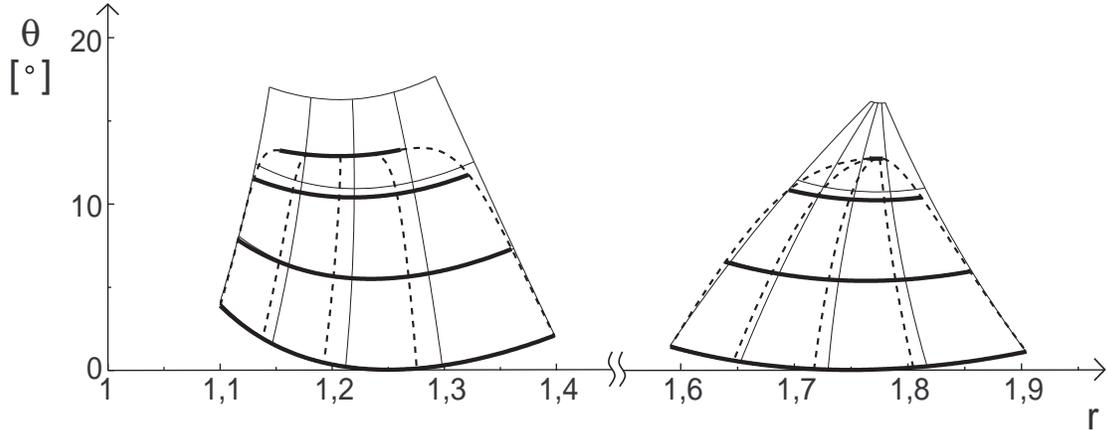}}
\caption{\small Nondegenerate evolution of the brane}
\label{fig:evol1}
\end{figure}

To demonstrate the effect of nondegeneracy on the evolution of the
brane, let us solve the approximate equation
\begin{equation}
\Psi_{app} \equiv \tt^2 - f^2 \t'^2 - h + \a \t^q = 0.
\label{eq:eom2}
\end{equation}
For that purpose, we must add a term proportional to $\t^q$ to the
Hamiltonian of section \ref{sec:deg}, and modify the equations for
$\pi$ to $d_\lambda \pi = - \partial_x H - \pi \partial_\t H$
\cite{arn}. As a result, equations for $\pi_R$ and $\t$ as well as
the trivial equation for $\pi_t$ acquire new terms in comparison
to (\ref{eq:eqchart}),
$$\frac {d\pi_R}{dt} = \ldots - \frac{f\a' \t^q + q\a \t^{q - 1}
\pi_R}{2\pi_t}, \quad \frac {d\pi_t}{dt} = - \frac 12 q\a \t^{q -
1}, \quad \frac {d\t}{dt} = \ldots - \frac {\a \t^q}{\pi_t}.$$ The
initial conditions stay unchanged, except for the condition for
$\pi_t$ which now reads $\pi_t = - \sqrt{h_* + \hat \t_*^2 - \a_*
\t_*^q}$. For numerical calculations, it is convenient to write
equation for $\t$ in terms of momenta, as $d_t\t = - \pi_R^2/\pi_t
+ \pi_t$. This equation conserves the value $H = 0$ better than
the equation with $\a$-term cited above, since it leads to $d_t H
= 0$ rather than $d_t H \propto H$.  In figure \ref{fig:evol1}
we depict two 3-branes living in a 4-dimensional space with the
same properties as those in figure \ref{fig:evol}, but with such a
large $\a$ that $\tt$ at the initial moment is suppressed by the
factor 1/10. For comparison, we added branes with zero $\a$,
depicted along with their characteristics by the light lines. The
nondegenerate evolution is considerably slower than the degenerate
one at the beginning, but at the time about 0.2 before
reconnection the two modes of evolution become practically
indistinguishable.

\section{Conclusion} \label{sec:con}

We have investigated how a piece of brane wrapped around a black
hole separates from the rest after the black hole is knocked out
into an extra dimension. The process is of interest since it
reflects nontrivial features of brane dynamics, and can affect
signatures of mini black holes produced in high energy collisions.
We have shown that the worldsheet becomes isotropic at the
reconnection point, and found how fast the brane approaches the
degenerate mode of evolution, with the worldsheet isotropic
everywhere, as it comes close to that point. The rate of
relaxation depends on the dimension of the brane: the higher the
dimension, the faster the brane becomes isotropic. The relaxation
is fast enough to guarantee that the two parameters characterizing
the motion of the brane just before reconnection which we have
computed for an isotropic brane, the angle at which the brane
moves and the rate of change of its curvature, do not change when
we pass to a timelike brane.

The main idea of our approach was that, instead of solving the
equation of motion from the very start, we restricted ourselves to
the last moments before reconnection. In this way we were able to
describe the behavior of the brane at this stage of its evolution,
but we had to resign on determining the characteristics of the
process as a whole. In particular, we could not compute the total
time it takes the black hole to separate from the brane.

In the simplified description we have adopted from \cite{fla}, the
brane passes through itself at the moment of reconnection and
develops an expanding intermediate domain between the piece
attached to the black hole and the bulk part where the black hole
resided before. To correct this picture, we must assign the brane
a finite thickness, while keeping it devoid of self-gravitation.
This can be most easily done by identifying the brane with a
domain wall composed of a scalar field with double-well potential.
In \cite{fla2} it was shown, by approximating the angular
dependence of the scalar field with the help of Chebyshev
polynomials, that a thick brane truly splits up at reconnection.
Instead of solving the equation for the scalar field, one can pass
to an effective theory in which the smearing of the scalar field
over a finite domain is taken into account by including curvature
corrections into the Lagrangian of the brane. This approach was
used to study static brane-black hole systems in \cite{czi}.
However, the resulting formulas are extremely lengthy even without
the time derivatives; thus, to extend our investigation to a thick
brane, a more promising approach seems to be that of ref.
\cite{fla2}.

\vskip 5mm \noindent {\it Acknowledgement.} This work was
supported by the grant VEGA 1/1008/09.

\appendix
\section{Two-step expansion} \label{2step}

After expanding $\F$ in $\t$, we can expand $\t$ in $\Dr$ and
$\Dt$ and solve the resulting infinite ``tower'' of algebraic
equations order by order. To see how the method works, consider
the nonphysical case $q = 1$. Equation for $\t$ reads
$$\tt^2 - f^2 \t'^2 - h = - \a \t - \b \t^2 - \g \t^3 - \ldots,$$
where $\b$ and $\g$ are of the form (only terms with the highest
order derivatives of $\t$ are listed)
$$\b = \frac {2\a}{\cal H} (\ttt - f^2 \t'') + \ldots, \quad \g = \frac
1{2{\cal H}} (-\tt \bt + f^2 \t' \b') + \ldots = - \frac \a{{\cal
H}^2} [\tt \dddot \t - f^2 (\tt \ddot \t' + \t' \tt'') + f^4 \t'
\t'''] + \ldots$$ Expand $\t$ into the series (\ref{eq:expth}) and
consider terms of order $(m, n)$ in the equation written above; by
definition, these are the terms appearing in front of $\Dt^m
\Dr^n$ with the factor $1/(m!n!)$ suppressed. On the left hand
side we have an expression of the form
$$2\t_{10} \t_{m + 1,n} + \mbox{terms with } \t_{lower}^2 -
h_n \delta_{m0},$$ where $\t_{lower}$ are $\t$'s with the sum of
the indices not exceeding $m + n$. In this expression, all
$\t_{lower}$ have the first index not exceeding $m$ and all but
one $\t_{lower}$ have the second index not exceeding $n$. The
exception is $\t$ with the indices $(m - 1,n + 1)$ appearing in
the term $- 2m f_0^2 \t_{11} \t_{m - 1,n + 1}$. Suppose for a
moment that all right hand sides are zero. Then we can find all
$\t$'s up to the order $N$ by solving the equations ``row by
row'', first the equations $(0, 0)$, $(0, 1)$, $\ldots$, $(0, N -
1)$, then the equations $(1, 0)$, $(1, 1)$, $\ldots$, $(1, N -
2)$, and so forth. In this way we have obtained the formulas
(\ref{eq:coeff}) in section \ref{sec:deg}. Let us now return to
the equations with nonzero right hand sides. The $\a$-term
contributes only ``safe'' $\t$'s with the first index not
exceeding $m$ and the second index not exceeding $n$, but the
remaining terms are producing also ``dangerous'' $\t$'s with one
or another index exceeding the corresponding limit value. Denote
the coefficients of expansion of a function $u (r, t)$ into the
powers of $\Dr$ and $\Dt$ by $u_{mn}$ and the coefficients of
expansion of a function $v (r)$ into the powers of $\Dr$ by $v_m$.
The ``dangerous'' $\t$'s can be identified from the structure of
the terms with the highest order derivatives of $\t$ in $\b$,
$\g$, $\ldots$: $\t_{\mu \nu}$ with $\mu = m + 1$, $m + 2$, $m +
3$, $\ldots$ and maximum $\nu$ appear in $\b_{m - 1, n - 2}
(\t^2)_{12} \sim \t_{m + 1, n - 2} \t_{10} \t_{02}$, $\b_{m, n -
4} (\t^2)_{04} \sim \t_{m + 2, n - 4} \t_{02}^2$, $\g_{m, n - 6}
(\t^3)_{06} \sim \t_{10} \t_{m + 3, n - 6} \t_{02}^3$, $\ldots$,
and $\t_{\mu \nu}$ with $\nu = n + 1$, $n + 2$, $n + 3$, $\ldots$
and maximum $\mu$ appear in $\b_{m - 2, n - 1} (\t^2)_{21} \sim
\t_{m - 2, n + 1} \t_{10} \t_{11}$, $\b_{m - 2, n} (\t^2)_{20}
\sim \t_{m - 2, n + 2} \t_{10}^2$, $\g_{m - 3, n} (\t^3)_{30} \sim
\t_{11} \t_{m - 4, n + 3} \t_{10}^3$, $\ldots$ In this way we find
that the ``dangerous'' $\t$'s are located inside the triangle $\mu
> m$, $\nu \le n - 2(\mu - m)$, and the trapezoid $\nu > n$, $\mu
\le m - 2$ for $\nu = n + 1$ and $\mu \le m - 2(\nu - n - 1)$ for
$\nu > n + 1$. For $q > 1$ both domains flatten, the higher $q$
the more. The crucial observation is that the domains are embedded
into the triangle $\mu \ge 0$, $\nu \ge 0$, $\mu + \nu \le m + n$;
in other words, all ``dangerous'' $\t$'s are of the type
$\t_{lower}$. Thanks to that we can solve the equations after
grouping them appropriately, this time ``triangle after
triangle'': first the equation $(0, 0)$, then the equations $(0,
1)$, $(1, 0)$, then the equations $(0, 2)$, $(1, 1)$, $(2, 0)$,
and so forth.

The expansion in $\Dr$ and $\Dt$ can be also used to determine the
free function $\a$ in the expansion of $\F$ from the known
function $\F_0$, or equivalently, $\xi_0$. From the formula $\xi =
\a + \b \t + \g \t^2 + \ldots$ we find $\xi_{0n} = \a_n + \b_{0, n
- 2} \t_{02} + \ldots + \g_{0, n - 4} \t_{02}^2 + \ldots$, so that
for the relation we seek we need to know the maximum order of
$\a$'s entering $\b$'s, $\g$'s etc. with the first index zero. We
must keep in mind, however, that $\a$'s appear in $\b$'s, $\g$'s
etc. also implicitly through $\t$'s, since all $\t_{kl}$ with $k >
0$ can be expressed in terms of $\t_{0j}$ and $\a_j$ by solving
the corresponding algebraic equations order by order. Suppose
again that $q = 1$. The coefficient $\xi_{0n}$ first appears in
the equation for $\t_{2n}$, which is of the form $2\t_{10} \t_{2n}
+$ terms with $\t_{lower}^2 = - \xi_{0n} \t_{10}$; hence all
$\t_{kl}$ contained in $\xi_{0n}$ are from the triangle $k + l \le
n + 1$. To identify $\a$'s entering $\xi_{0n}$, it suffices to
consider $\a$'s that are present {\it explicitly} in $\t$'s from
the triangle, after $\b$'s, $\g$'s etc. are expressed in terms of
$\a$'s and $\t$'s (which are also from the triangle). Let us write
equation for $\t_{kl}$ as
$$2\t_{10} \t_{kl}  + \mbox{terms with } \t_{lower}^2 -
h_l \delta_{k1} = - \a_l \t_{k - 1, 0} - \ldots - \b_{k - 3, l}
\t_{10}^2 - \ldots - \g_{k - 4, l} \t_{10}^3 - \ldots$$ The
enlisted expansion coefficients of $\b$, $\g$ etc. contain the
highest order $\a$'s for high enough $k$. They appear in $\b_{k -
3, l} \sim (\t')_{k - 3, 0} (\a')_l \propto \a_{l + 1}$ for $k >
3$ ($l < n - 2$), in $\g_{k - 4, l} \sim (\t')_{k - 5, 0}
(\t')_{10} (\a'')_l$ $\propto \a_{l + 2}$ for $k > 5$ ($l < n -
4$) etc. (Other coefficients contribute at best $\a$'s of the same
order. For example, for $3 < k \le 5$ we have $\g_{k - 4,l} \sim
(\tt \bt)_{k - 4,l} \sim \t_{10} \b_{k - 3,l} \propto \a_{l +
1}$.) All these $\a$'s are clearly of lower order than $\a_n$.
Furthermore, the highest order $\a$ appears in $\a_{l - 2}
\t_{02}$ for $k = 1$ ($l \le n$), in $\a_l \t_{10}$ for $k = 2$
($l \le n - 1$) and in $\a_l \t_{20}$ and $\b_{0l} \t_{10}^2$,
$\b_{0l} \sim (\t')_{01} (\a')_{l - 1} \propto \a_l$, for $k = 3$
($l \le n - 2$). These are again $\a$'s of lower order than
$\a_n$. If we rise $q$ and use the same procedure, the triangle
will be larger but the relative order of $\a$'s will be lower, so
that the net result will be the same. Thus, for any $q$ we have
$\xi_{0n} = \a_n +$ terms with $\a_{lower}$; and by solving these
equations order by order with respect to $\a_n$ we obtain $\a_n =
\xi_{0n} +$ terms with $\xi_{lower}$.

\section{One-step expansion} \label{1step}
\setcounter{equation}{0}
\renewcommand{\theequation}{B-\arabic{equation}}

Equations $\P = 0$ and $\Psi = 0$ expanded in $\Dr$ and $\Dt$
transform into two ``towers'' of algebraic equations $\P_{mn} = 0$
and $\Psi_{mn} = 0$ for the coefficients $\F_{mn}$ and $\t_{mn}$.
Let us look at the structure of these equations. Rewrite $\P$ as
\begin{equation}
\P = \xi \Ft - \eta \F' - 2\zeta \F - qh\F, \label{eq:P1}
\end{equation}
where
\begin{equation}
\xi = \ta\ \tt, \quad \eta = \ta\ f^2 \t', \quad \zeta = \ta\ \T.
\label{eq:Pcoeff}
\end{equation}
We can see from these formulas that the equations $\Phi_{mn} = 0$
of $N$th order (with $m + n = N$) contain only $\F_{kl}$ of $N$th
or lower orders (with $k + l \le N$). Indeed, because of the
factor $\ta$ in the definitions of $\xi$, $\eta$ and $\zeta$, the
zeroth order term is missing in the expansions of all three
functions into the powers of $\Dt$ and $\Dr$, therefore the
coefficients $\F_{kl}$ contributed to $\Phi_{mn}$ by the first
three terms in $\Phi$ are of maximum order $N$, $N$ and $N - 1$
respectively. In fact, the expansion of $\eta$ starts with the
terms of {\it second} order, namely with $\eta_{20}$ and
$\eta_{11}$, hence the second term in $\Phi$ yields coefficients
$\F_{kl}$ of maximum order $N - 1$, too. Explicitly,
$$\P_{mn} = (m \xi_{10} - q h_0) \F_{mn} + \mbox{terms proportional
to} \ \F_{lower}.$$ For $\xi_{10}$ we have
$$\xi_{10} = [\t \tt]_{10} = [(\t_{10} \Dt + \ldots) (\t_{10}
 + \ldots)]_{10} = \t_{10}^2 = h_0,$$
(the last equality follows from $\F_{00} = 0$, which is now a
consequence of $\P_{00} = 0$), hence
\begin{equation}
\P_{mn} = (m - q) h_0 \F_{mn} + \mbox{terms proportional to} \
\F_{lower}. \label{eq:Pmn}
\end{equation}

From the expression for $\Phi_{mn}$ it is evident that if $q$ is
not a natural number and $\Phi_{mn}$ vanishes for all $m$ and $n$,
$\F_{mn}$ must vanish for all $m$ and $n$, too. Indeed, if the
factor in front of $\F_{mn}$ is nonzero, equation $\Phi_{mn} = 0$
determines the coefficient $\F_{mn}$ in terms of coefficients of
lower order; and if {\it all} factors are nonzero, which is the
case for any non-natural $q$, these lower order coefficients have
been already fixed to zero by previous equations. We already know
that the true solution to equation $\Phi = 0$ merge with the
degenerate one at the point of reconnection. Now we can see that
for all non-natural values of $q$ a stronger statement is valid:
the only solutions that are regular at the point of reconnection
(can be Taylor-expanded in $\Dr$ and $\Dt$) are those that are
{\it exactly} degenerate.

The ``regularity means degeneracy'' claim for non-natural $q$'s
can be obtained immediately from the fact that the expansion of
$\F$ in $\t$ starts from $\t^q$. Since $\F$ is by definition
quadratic in the first derivatives of $\t$, and since it expands
into non-natural powers of $\t$, the function $\t$ itself must
expand into non-natural powers of $\Dr$ and $\Dt$ unless $\F$ is
identically zero.

Let us now proceed to the case when $q$ is natural, $q = 1, 2, 3,
\ldots$ (For completeness, we consider also odd values of $q$,
although only even values are physically relevant.) Everything
works as with non-natural $q$'s until we arrive at the equation
$\P_{q0} = 0$. The equation is satisfied identically, therefore
the coefficient $\F_{q0}$ is {\it at this stage} free. However, it
can be fixed to zero at the next step, by the equation $\P_{q1} =
0$; or if not, then at the next-to-next step, by the equation
$\P_{q2} = 0$; and so forth. The expressions $\Phi_{qj}$ on the
left hand side of these equations can be transformed into linear
combinations of $j$ coefficients $\F_{q0}$, $\F_{q1}$, $\ldots$,
$\F_{q, j - 1}$, because other coefficients entering them are
either zero or can be expressed, by using ``their'' equations, as
linear combinations of the coefficients we have listed. Thus, we
have
\begin{equation}
\P_{qj} = C_{1j} \F_{q,j - 1} + C_{2j} \F_{q, j - 2} + \ldots +
C_{jj} \F_{q0}. \label{eq:Pqj}
\end{equation}
Coefficients $C_{ij}$ appearing here can be written as
\begin{equation}
C_{ij} = \mbox{ terms of the form } \t^2,\ \t^4, \ldots - q \binom
ji h_i. \label{eq:Cij}
\end{equation}
where by ``terms of the form $\t^2$, $\t^4$, $\ldots$'' we
understand terms proportional to the product of the corresponding
number of coefficients $\t_{mn}$. Some of these terms appear in
front of $\F_{q,j - i}$ in the original expression for $\P_{qj}$,
and the rest are secondary terms contributed by $\F$'s of higher
order. In the primary terms, higher powers of $\t$ come from the
expansion of $\ta$, while in the secondary terms they are produced
also by inserting for higher order $\F$'s the expressions obtained
from ``their'' equations. Indeed, by putting the expression
(\ref{eq:Pmn}) equal to zero we obtain $\F_{mn} \sim \mbox{ terms
of the form } \t^2,\ \t^4, \ldots \times \F_{lower}/h_0$ for any
$m \ne q$, therefore each consecutive order of $\F_{mn}$ brings
with itself at least one extra factor $\t^2$ into $C_{ij}$.

The next step is to compute $\t$'s from equations $\Psi_{mn} = 0$.
The equations can be written as
$$2\t_{10} \t_{m + 1,n} + \mbox{terms with }
\t_{lower}^2 - h_n \delta_{m0} = - \F_{mn},$$ and they yield, when
solved order by order, expressions for $\t_{mn}$ containing
$\t_{0j}$ and $\F_{qj}$ only. The expressions can be divided into
``degenerate'' part $\t^{(0)}_{mn}$ that does not contain
$\F_{qj}$, and optional extra terms proportional to the powers of
$\F_{qj}$. First few $\t^{(0)}$'s have been computed at the
beginning of section \ref{sec:deg}. After the expressions for
$\t$'s are inserted into the expressions for $C$'s, they also
split into two parts, ``degenerate'' part $C^{(0)}_{ij}$ and
optional $\F$-terms. Extra terms first appear in $C_{2q,2q}$. They
include, in particular, $\F$-term coming from the primary term $q
\xi_{1,2q} \F_{q0}$ in $\P_{q, 2q}$, with $\xi_{1,2q} = 2 \t_{10}
\t_{1,2q} +$ terms of the form $\t_{lower}^2$, $\t_{lower}^4$,
$\ldots$ The term is proportional to $\F_{q0}$, since $\t_{1,2q}$
can be expressed in terms of $\F_{0,2q}$, which in turn can be
expressed in terms of $\F_{q0}$ (most simply by using the relation
$\F \sim \a \t^q$). All other extra terms in $C_{2q,2q}$ are
proportional to $\F_{q0}$, too.

For $C^{(0)}_{ij}$, as well as for the coefficients at the powers
of $\F_{qj}$, if present, we have expressions whose complexity
grows progressively with their order. However, after computing
them explicitly for the first few coefficients $C_{ij}$ we have
found that the terms comprising them mutually cancel. We have
checked this, partly with the help of MAPLE, for $C_{ij}$ with the
first index running up to $i = 4$. Thus, the first few equations
$\P_{qj} = 0$ reduce to $0 = 0$. The considerations of section
\ref{sec:true} suggest that this is true for {\it all} equations
$\P_{qj} = 0$. In other words, for natural $q$'s there holds a
direct opposite to what we have established for non-natural $q$'s:
regularity does {\it not} mean degeneracy. On the contrary, it
means maximum nondegeneracy in the sense that all $C_{ij}$ are
identically zero and all $\F_{qj}$ are free.

\end{document}